\documentstyle[epsfig]{mn}
\begin{document}

\title[] {
Uses and limitations of relativistic jet proper motions:
lessons from galactic microquasars
}
\author[R.~P.~Fender] {R. P. Fender \\
Astronomical Institute `Anton Pannekoek', University of Amsterdam, 
Kruislaan 403, 1098 SJ Amsterdam, The Netherlands\\ }

\maketitle

\begin{abstract}

It is shown that the two-sided jet proper motions observed from the
galactic microquasars GRS 1915+105 and GRO J1655-40 in practice only
allow us to place {\em lower limits} on the Lorentz factors of the
outflows. As a consequence, it is not possible to rule out the
possibility that jets from X-ray binaries are just as relativistic as
those from active galactic nuclei (AGN).  This results from the fact
that distance estimates place the sources, within uncertainties, at
the maximum distance $d_{\rm max}$ which corresponds to an intrinsic
velocity $v=c$. The general case is explored, for a range of intrinsic
Lorentz factors and angles to the line of sight, and it is shown that
a source of significantly relativistic jets will nearly always be
observed close to $d_{\rm max}$ and as a result it is unlikely that we
will ever be able to measure with any accuracy the Lorentz factor of a
jet from two-sided proper motions. We will generally not be able to do
more than place a lower limit on the Lorentz factor of the flow, and
this limit is naturally even lower in the cases where we only observe
the approaching jet.  On the other hand, under the assumption that any
two-sided jets we see are intrinsically relativistic, we can
confidently place the source at a distance $d \sim d_{\rm max}$. As a
result, observations of two-sided proper motions in relativistic jets
from AGN would be extremely important for calibration of the
cosmological distance scale.  While the proper motions do not allow us
the measure the Doppler shifts associated with the jets, the ratio of
proper motions will correspond to the ratio of frequencies of any
emission lines emitted by both jets, which will aid in searching for
such lines. Furthermore, it is shown that if the jet is precessing,
the product of the proper motions as a function of angle to the line
of sight may be used to determine if the jet is only mildly
relativistic.

\end{abstract}



\section*{Introduction}

\begin{figure*}
\centerline{{\epsfig{file=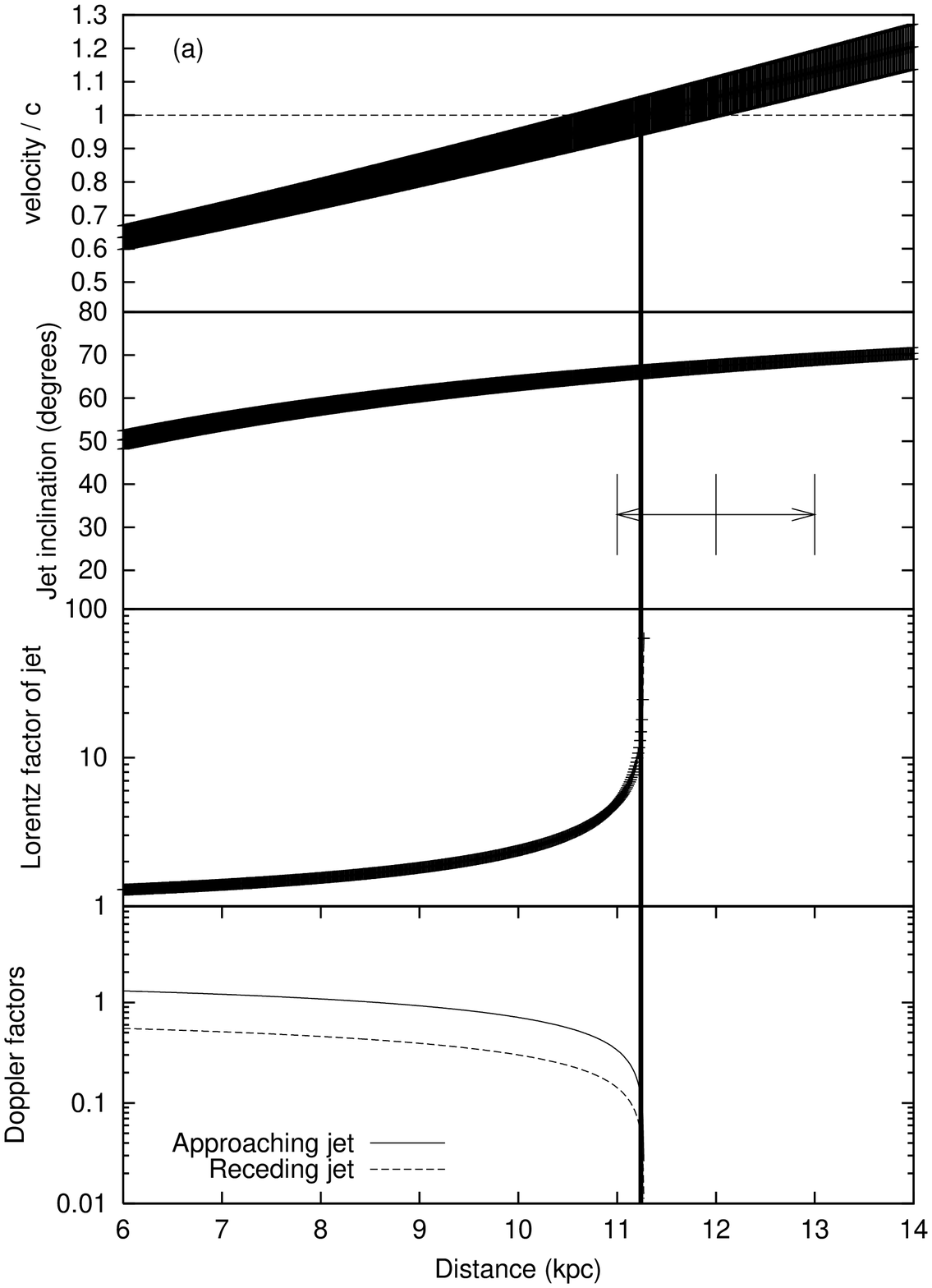, angle=0, width=8cm}}\quad{\epsfig{file=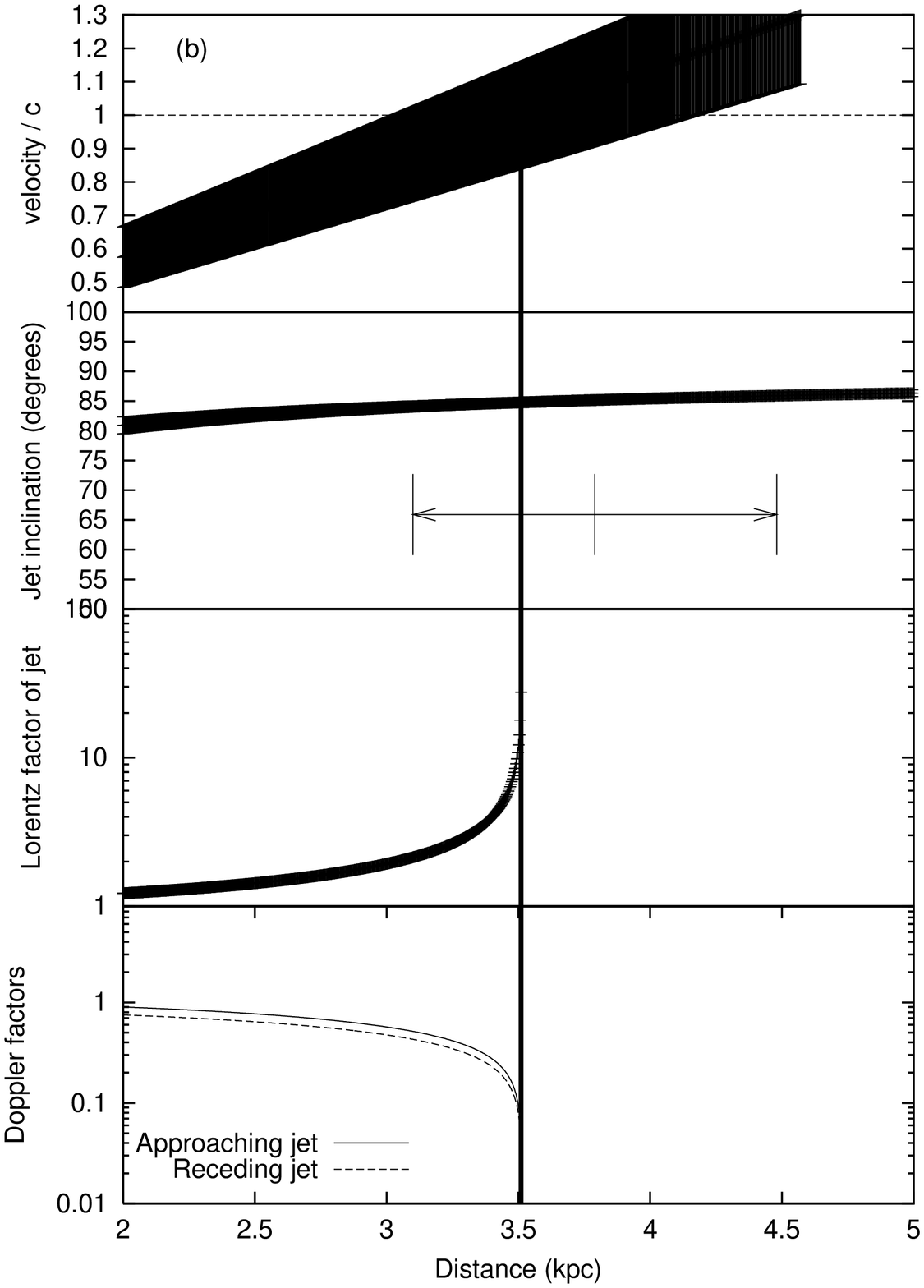, angle=0, width=8cm}}}
\caption{Solutions to $\beta$ and $\theta$, and the resultant values
for the Lorentz ($\Gamma$) and Doppler ($\delta$) factors, for GRS 1915+105 (a, left) and GRO J1655-40
(b, right). The solution for GRS 1915+105 are based upon the proper
motions reported in Fender et al. (1999), and those for GRO J1655-40
from Hjellming \& Rupen (1995). Distance estimates, from Dhawan et
al. (2000) and Greene et al. (2002), are  indicated in the second panel.
In both cases the distance estimates
include $d=d_{\rm max}$, for which $\Gamma$ is infinitely large and
the Doppler factors infinitely small. Therefore, the observed
two-sided proper motions have not constrained the Lorentz factors of
these jets.}
\end{figure*}

Relativistic jets, outflows of matter from regions close to accreting
black holes and neutron stars, remain amongst the most spectacular yet
poorly explained phenomena in high-energy astrophysics. They are
ubiquitous amongst Active Galactic Nuclei (AGN) powered by
supermassive black holes, as well in spectrally hard and transient
outbursting states of stellar mass black holes and neutron stars in
X-ray binary systems (XRBs) -- see e.g. Hughes (1991); Mirabel \&
Rodriguez (1999).

One of the key questions in the study of these jets is `how
relativistic are they' -- i.e. what is the Lorentz factor ($\Gamma =
(1-\beta^2)^{-1/2}$, where velocity $v=\beta c$) of the flow ? In AGN,
the highest inferred Lorentz factors are $\sim 30 h^{-1}$ ($h = H_0 /
100$ km s$^{-1}$ Mpc$^{-1}$, where $H_0$ is the Hubble constant) --
see e.g. Vermeulen \& Cohen (1994); Jorstad et al. (2001). Multiple
recent detections of arcsec-scale X-ray jets from AGN with {\em
Chandra} (e.g. Harris \& Krawczynski 2002) indicate that relativistic
flow velocities are maintained over large distances from the jet base.

In two XRBs we have a clear advantage over studies of AGN, in that we
can observe the proper motions and flux ratios of both approaching and
receding radio knots (a.k.a. `blobs', `plasmons' etc.)  associated
with the same ejection event. Observations of ejections from these two
sources, GRS 1915+105 (Mirabel \& Rodriguez 1994; Fender et al. 1999;
Rodriguez \& Mirabel 1999; Fender et al. 2002) and GRO J1655-40
(Tingay et al. 1995; Hjellming \& Rupen 1995), provide us with unique
diagnostics of the jet geometry. However, as shall be discussed in
this paper, and contrary to widespread misconception, they have {\em
not} allowed us to measure the Lorentz factor of the flow in either
case.

\section*{XRB Jet Proper motions}

\begin{table*}
\begin{tabular}{ccccccc}
& $\mu_{\rm app} (mas/d) $ & $\mu_{\rm rec}$ (mas/d) & $\beta \cos \theta$ &
$\theta_{\rm max}$ (degrees) & $d_{\rm max} (kpc)$ & REF \\ 
\hline
GRS 1915+105 & $17.6 \pm 0.4$ & $9.0 \pm 0.1$ & $0.323 \pm 0.016$ &
71 & 13.7 & MR94\\
             & $23.6 \pm 0.5$ & $10.0 \pm 0.5$ & $0.41 \pm 0.02$ & 66
& 11.2 & F99\\
\hline
GRO J1655-40 & 54 & 45 & 0.09 & 85 & 3.5 & HR95\\
\hline
\end{tabular}
\caption{Simultaneous measurements of approaching and receding knot
velocities in the jets from two galactic black hole binaries. Refs:
MR94 = Mirabel \& Rodriguez 1994; F99 = Fender et al. 1999; HR95 =
Hjellming \& Rupen 1995. HR95 do not provide estimates of their
measurement uncertainties.}
\end{table*}

In the following discussion we will consider relativistic jets in
which proper motions associated with approaching ($\mu_{\rm app}$) and
receding ($\mu_{\rm rec}$) components can be measured (assuming that
both sides of the jet have been correctly associated with the same
ejection event, the higher proper motion of the two always correponds
to $\mu_{\rm app}$). A {\em key} point in the following discussion is
the assumption of symmetry in ejection velocity for both sides of the
jet; possible exceptions to this will be discussed at the end.

As described in Mirabel \& Rodriguez (1994), measurement of $\mu_{\rm
app}$ and $\mu_{\rm rec}$ allows a determination of the following
product:

\[
\beta \cos \theta = \frac{(\mu_{\rm app}-\mu_{\rm rec})}{(\mu_{\rm
app}+\mu_{\rm rec})}
\]

where $\theta$ is the angle of the ejection to the line of sight and
$\mu_{\rm app}$, $\mu_{\rm rec}$ are the approaching and receding
proper motions respectively (see also Rees 1966; Blandford, McKee \&
Rees 1977). 

Once the proper motions are measured, the angle of ejection, $\theta$,
and consequently the intrinsic velocity, $\beta$, are uniquely
determined for every distance since

\[
\tan \theta = \frac{2d}{c}\left(\frac{\mu_{\rm app}\mu_{\rm
rec}}{\mu_{\rm app}-\mu_{\rm rec}}\right)
\]

\noindent and the product $\beta \cos \theta$ is already known.

The variation of $\beta$ and $\theta$ as a function of distance for GRS
1915+105 was presented in Fender et al. (1999). There is a maximum
distance to the source corresponding to $\beta = 1$ (i.e. $\Gamma =
\infty$):

\[
d_{\rm max} = \frac{c}{\sqrt(\mu_{\rm app}\mu_{\rm rec})}
\]

At this upper limit to the distance you also find the maximum
angle of the jet to the line of sight,

\[
\theta_{\rm max} = \cos^{-1} \frac{(\mu_{\rm app}-\mu_{\rm rec})}{(\mu_{\rm
app}+\mu_{\rm rec})}
\]

\begin{figure*}
\centerline{{\epsfig{file=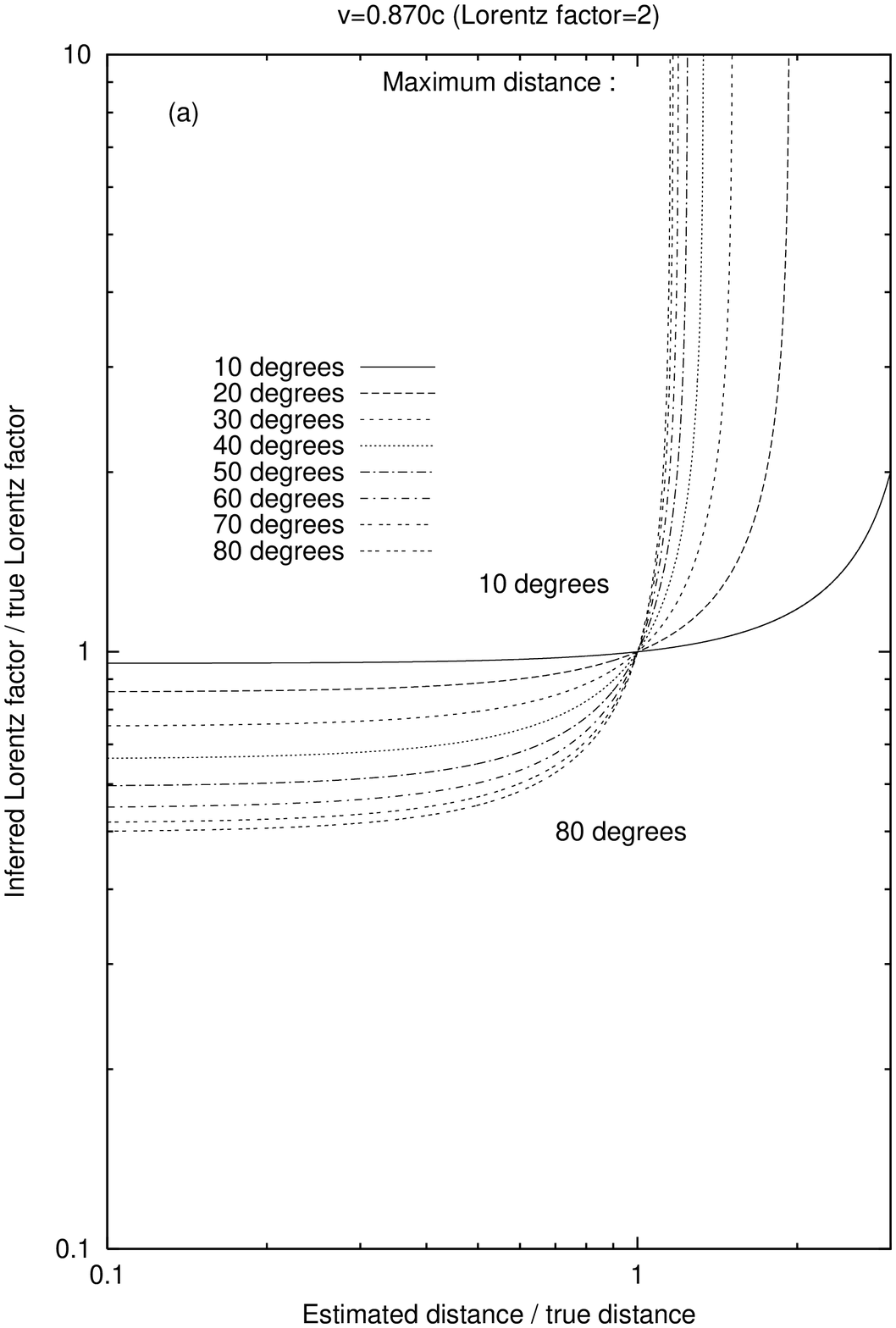,
angle=0,width=8cm,height=10cm}}\quad{\epsfig{file=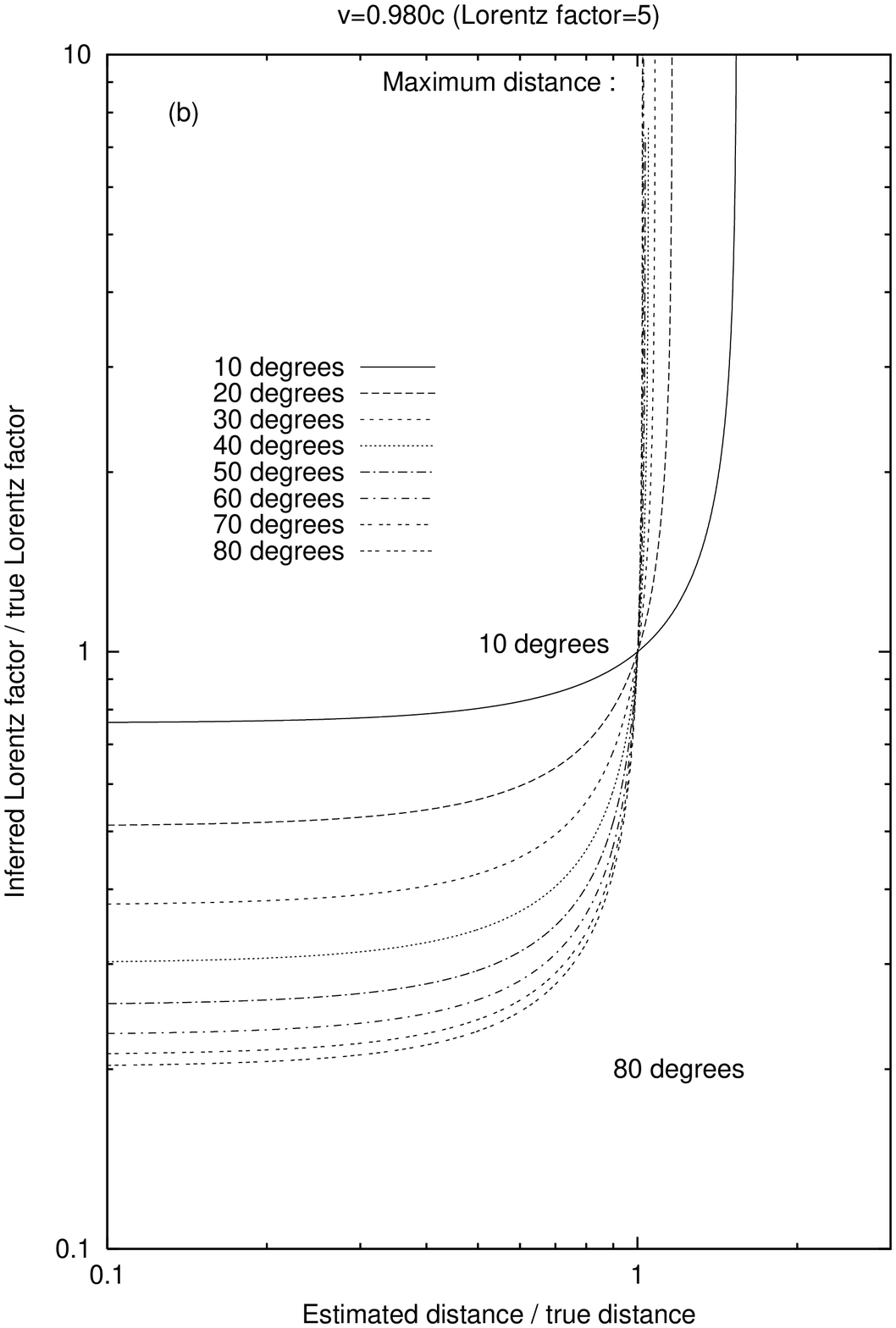, angle=0,
width=8cm, height=10cm}}}
\centerline{{\epsfig{file=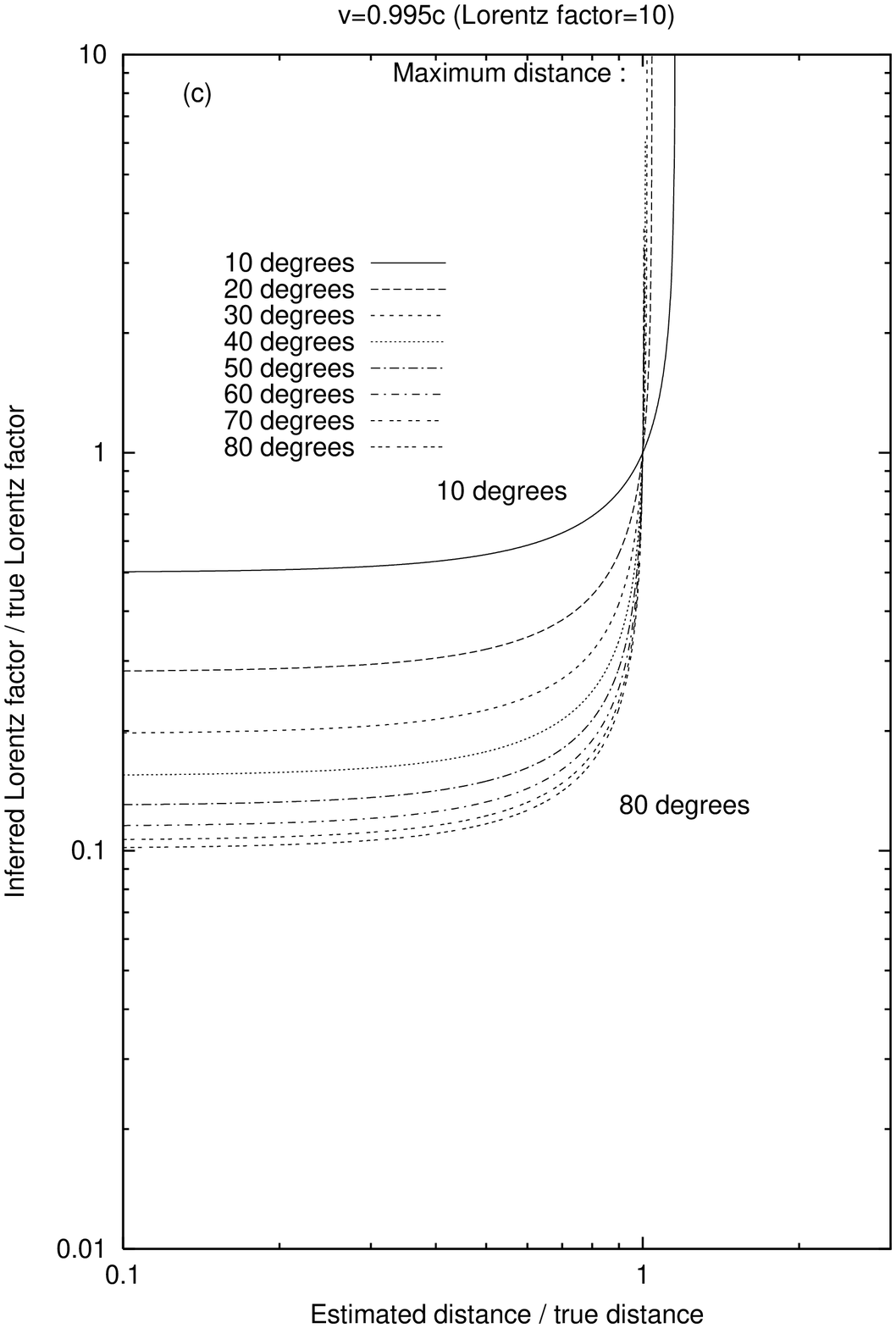, angle=0,
width=8cm,height=10cm}}\quad{\epsfig{file=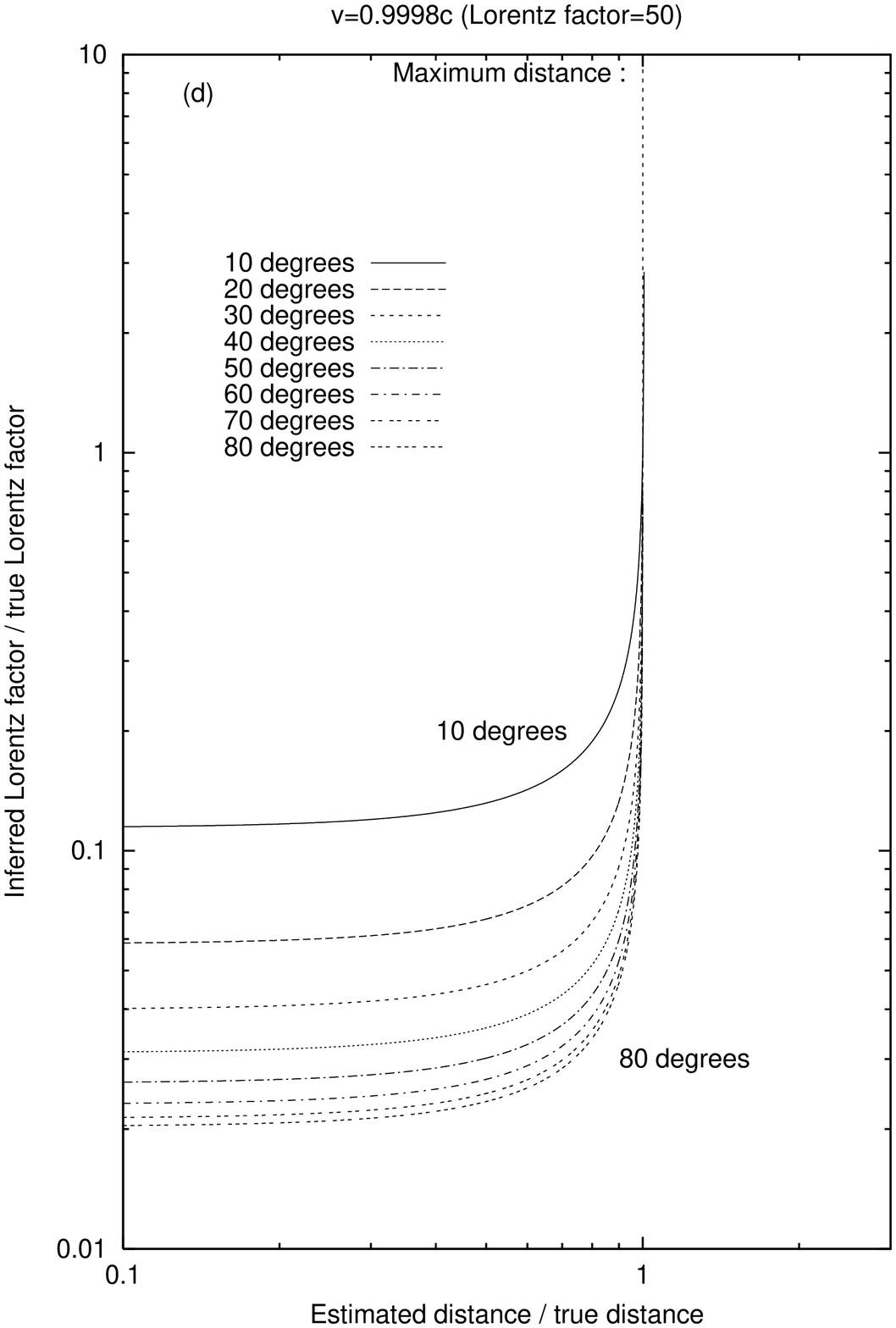, angle=0, width=8cm, height=10cm}}}
\caption{Indications of the uncertainties in derived Lorentz factors
for jets with different intrinsic velocities and inclinations. Each
panel shows the Lorentz factor which would be derived for the jets,
given a distance estimate expressed as a fraction of the real
distance, for a range of jet angles. The four panels show the
solutions for four different intrinsic Lorentz factors.  The points at
which the functions reach the top of the panel correspond to $d_{\rm
max}$ for that angle and Lorentz factor.  For all solutions except
those with the lowest velocities and smallest angles, the true
distance lies very close to $d_{\rm max}$, indicating (a) it will be
practically impossible to constrain $\Gamma$, (b) sources so observed
will in reality lie close to $d_{\rm max}$, allowing a distance
estimate based on the proper motions alone (see also Fig 3). Note that
the lower two panels have different ordinates than the upper two
panels.  }
\end{figure*}

\section*{Microquasar measurements}

For two galactic X-ray binary systems, GRS 1915+105 and GRO J1655-40,
$\mu_{\rm app}$ and $\mu_{\rm rec}$ have been measured for multiple
ejection events. The data are summarised in table 1.

Both sources have fairly accurate independent distance estimates. For
GRS 1915+105, Mirabel \& Rodriguez (1994) estimate a distance of $12.5
\pm 1.5$ kpc based on HI measurements. Dhawan, Goss \& Rodriguez
(2000) revise this distance estimate to $12 \pm 1$ kpc. The large
X-ray column and optical extinction to the source are in agreement
with this relatively large distance.

For GRO J1655-40, McKay \& Kesteven (1994) estimated a distance of 3.5
kpc, and Tingay et al. (1995) estimated a distance of 3--5 kpc. The
kinematic model fit performed by Hjellming and Rupen (1995) resulted
in a distance estimated of $3.2 \pm 0.2$ kpc.  Most recently, Greene,
Bailyn \& Orosz (2001) derive $d = 3.79 \pm 0.69$ kpc based on
modelling of optical data.

Comparison of these distance estimates with the values for $d_{\rm
max}$ listed in table 1 reveals immediately that all the distance
estimates place the sources very close to (or even beyond!) $d_{\rm
max}$. The result of this is that from such observations we can only
place a {\em lower limit} on the Lorentz factor of the jets.

This is illustrated in Figs 1(a),(b). In these figures the solutions
for $\beta$ and $\theta$, based on the observed proper motions, are
plotted as a function of distance to the sources. Also indicated are
the best distance estimates, as well as the Lorentz and relativistic
Doppler factors resulting from the solutions to $\beta$ and
$\theta$. What is clear, for both sources, is that the distance
estimates -- already fairly accurate -- cannot do more than place a
lower limit of 2--3 on the Lorentz factors of the ejections. No upper
limit is possible as the range of possible distances includes $d_{\rm
max}$. Consequently we can only place upper limits on the Doppler
shifts associated with the jets.

These figures clearly illustrate that for these two celebrated sources,
the measured proper motions combined with the distance uncertainties
do {\em not} allow us to measure how relativistic the jets are. In this
paper we shall show that this will almost always be the case.

\section*{Can we limit $\Gamma$ using the flux ratios?}

A further misunderstanding propagating in the literature is that the
flux ratio observed between the approaching and receding knots is
somehow an independent confirmation of any distance or velocity
measurement already derived from the proper motions. This asymmetry in
brightness between the approaching and receding knots is due to a
combination of classical Doppler and relativistic aberration effects, both
contained in the relativistic Doppler factor

\[
\delta = \frac{1}{\Gamma (1 - \beta \cos \theta)}
\]

An object moving at angle $\theta$ to the line of sight with velocity
$\beta$ (and resultant Lorentz factor $\Gamma$) will have an observed
surface brightness $\delta^k$ brighter, where $2 < k < 3$ ($k=2$
corresponds to the average of multiple events in e.g. a continuous
jet, $k=3$ corresponds to emission dominated by a singularly evolving
event). Therefore the ratio of flux densities from approaching and
receding knots -- measured at the same angular separation from the
core, so as to sample the knots at the same age in their evolution --
will be given by:

\[
\frac{S_{\rm app}}{S_{\rm rec}} = \left(\frac{\delta_{\rm
app}}{\delta_{\rm rec}}\right)^{k-\alpha}
\]

\noindent 

where $\alpha$ is the spectral index of the emitting region, defined
such that $S_{\nu} \propto \nu^{\alpha}$. This additional term
compensates for the Doppler shifted spectrum when observing at a
single frequency.  The ratio of the proper motions is simply the ratio
of the Doppler factors, so

\[
\frac{S_{\rm app}}{S_{\rm rec}} = \left(\frac{\mu_{\rm app}}{\mu_{\rm
rec}}\right)^{k-\alpha}
\]

Thus once $\mu_{\rm app}$ and $\mu_{\rm rec}$ have been measured, the
only additional information obtained by measuring the flux ratio
between approaching and receding jet relates to the parameter $k$.
Although it may seem counter-intuitive that the flux ratio should
remain constant as $\beta$ increases, this is because at the same time
$\theta$ is also increasing.  The meaning of $k$ will not be explored
in detail here; however a small point is worth making: in observations
in which we can be fairly confident that we have resolved a single
radio knot, then $k$ should have the value 3. If we measure a value
less than this it may indicate that the bulk velocity of the flow is
significantly less than the {\em pattern} velocity which we are
observing (for further discussion see e.g. Blandford et al. 1977).

\section*{The general case}

It is straighforward to calculate the proper motions for jets of a
given $\beta$ and $\theta$, and compare them to the values we would
derive using the method outlined above, for varying estimates of the
distance to the source.

In Figs 2(a--d) we plot the inferred Lorentz factor as a fraction of
the intrinsic Lorentz factor of the jet, as a function of the distance
estimated to the source expressed as a fraction of the true distance.
In each figure the different curves indicate different intrinsic
angles to the line of sight, and each of the four panels represents a
different intrinsic Lorentz factor (2, 5, 10, 50). The points at which
the curves intersect with the upper abscissa corresponds to $d_{\rm
max}$ for the particular combination of proper motions observed.

Apart from the smallest angles and lowest velocities, there is for all
the curves an extremely rapid variation in the inferred Lorentz factor
close to the true distance to the source. The figures demonstrate
clearly that it will be effectively impossible to measure the distance
accurately enough to constrain the Lorentz factor. A related point is
that all significantly relativistic jets will by necessity lie very
close to $d_{\rm max}$ (Fig 3). This leads to one useful conclusion --
if the jets we observe are intrinsically significantly relativistic,
which seems to be the case, then measurements of two-sided proper
motions will give us an accurate distance estimate. As a result, this
means that observations of two-sided jet proper motions in AGN, were
they ever to be observed, would be extremely useful for calibrating
the cosmological distance scale. Unfortunately, to date most
well-studied AGN are significantly Doppler-boosted (so-called `Doppler
favouritism'), implying small angles of the jets to the line of sight,
and resulting in no measurements of two-sided relativistic proper
motions so far.

As noted above, there is a glimmer of hope for the lowest velocities
and smallest angles, where the swing in the curves around the true
distance is not too dramatic. However (a) this discussion is really
concentrating on significantly relativistic jets, and (b) the smallest
angles to the line of sight will have the largest ratios of proper
motions and fluxes between the approaching and receding sides of the
jet, making the measurements increasingly hard to make.

\begin{figure}
\centerline{{\epsfig{file=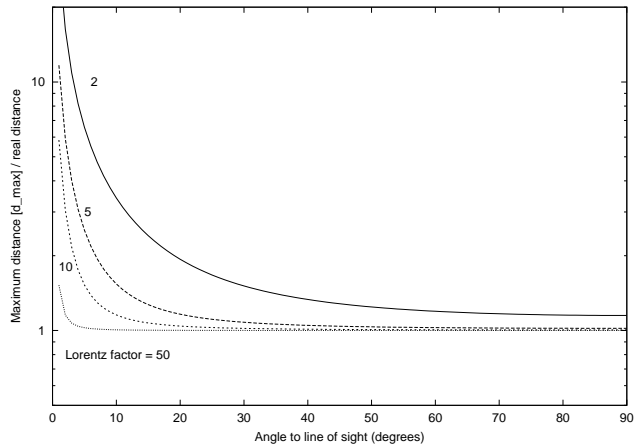, width=6cm, angle=270}}}
\caption{The ratio of $d_{\rm max}$ (see text) to the true source
distance, as a function of jet angle to the line of sight, for a range
of intrinsic Lorentz factors. Apart from the lowest velocities and the
smallest angles to the line of sight, relativistic jet sources will
always be observed at $d \sim d_{\rm max}$.
}
\end{figure}

This is illustrated in Fig 4, in which the proper motions (scaled to a
distance of 1 kpc) and resultant apparent velocity as a fraction of
the speed of light are plotted for different intrinsic Lorentz factors
as a function of angle to the line of sight. The receding proper
motions are very similar for all intrinsic Lorentz factors, but the
approaching proper motions are differing functions which peak at
progressively smaller angles (the peaks occur at $\theta \sim
1/\Gamma$ radians). Note that for both GRS 1915+105 and GRO J1655-40
the ratio of approaching to receding proper motions has been $<3$,
which, as this figure illustrates, indicates immediately that whatever
the Lorentz factor, they must be at large angles to the line of sight
(and therefore, consulting Fig 3, unless the jets are only mildly
relativistic, means that they must both lie at $d \sim d_{\rm max}$).
If we are hoping to measure the Lorentz factor from the proper motions
of a jet close to the line of sight then the ratio of proper motions
becomes increasingly large -- and therefore increasingly hard to
measure accurately. The ratio of fluxes is even greater, being the
ratio of proper motions raised to some power $k$ (at the same angular
separation), and so compared to the approaching component the receding
jet will appear to be extremely faint and slow moving. Most likely we
will observe only the approaching jet, or jet plus core if activity is
still ongoing (as has been the case to date for AGN).

\begin{figure}
\centerline{{\epsfig{file=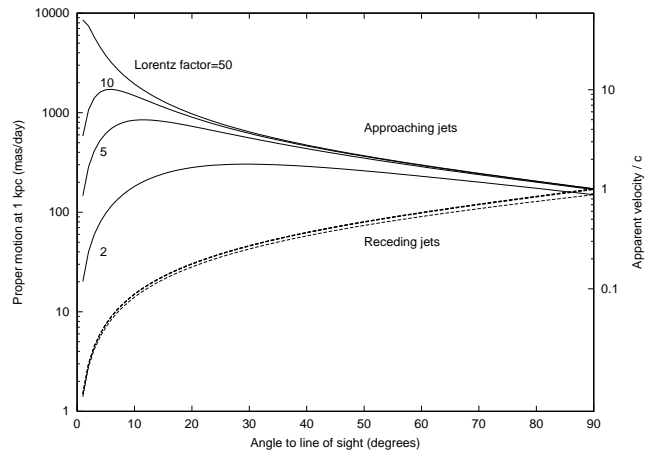, width=6cm, angle=270}}}
\caption{Proper motions for approaching and receding jets for the same
range in Lorentz factors, scaled to 1 kpc. The receding jets all show
approximately the same proper motions; the approaching jets peak at an
angle $\theta \sim 1/\Gamma$. Assuming this angle, an independent
lower limit on $\Gamma$ may be obtained.  At the smallest angles the
large difference in proper motions and fluxes (see text) will make
observations of two-sided jets close to the line of sight extremely
difficult. }
\end{figure}

This leads us to consider an alternative approach to at least limiting
the Lorentz factor. For a jet of apparent velocity $\beta_{\rm app}$,
the intrinsic Lorentz factor is at least as large as $\beta_{\rm
app}$, corresponding to the solution for $\theta = 1 / \Gamma$. In
this way observations of one-sided proper motions can allow us to
place a lower limit on the Lorentz factor. How accurate is this method
compared to two-sided proper motions ? In fact it can never place a
more constraining lower limit on $\Gamma$ than can be obtained by
measurement of two-sided proper motions.  This is natural, since the
lower limits to the Lorentz factors measured from one-sided proper
motions assume the jet is at its optimum angle, resulting in maximum
apparent velocity, which will generally not be the case.

\section*{What {\rm can} we learn ?}

We have established above that it will be practically impossible to do
more than place a lower limit on the Lorentz factor of a relativistic
jet from proper motions, whether one- or two-sided. However, the
proper motions themselves can be used to make a distance estimate to
the source, more accurate the more relativistic the jet intrinsically
is. What else can we learn from the proper motions ?

As already stated, the ratio of proper motions is also the ratio of
Doppler factors. This may be useful in associated unidentified lines
with a jet, even though the absolute value of the Doppler shift cannot
be predicted from the proper motions. For example, to check if an
unidentified feature at wavelength $\lambda_a$ is a Doppler shifted line of
from the approaching jet, look at wavelength $\lambda_r = \lambda_a
\times \mu_{\rm app} / \mu_{\rm rec}$ to try and find the line from
the receding jet (obviously more difficult since it will be Doppler
de-boosted).

\subsection*{Mildly relativistic precessing jets}

\begin{figure}
\centerline{{\epsfig{file=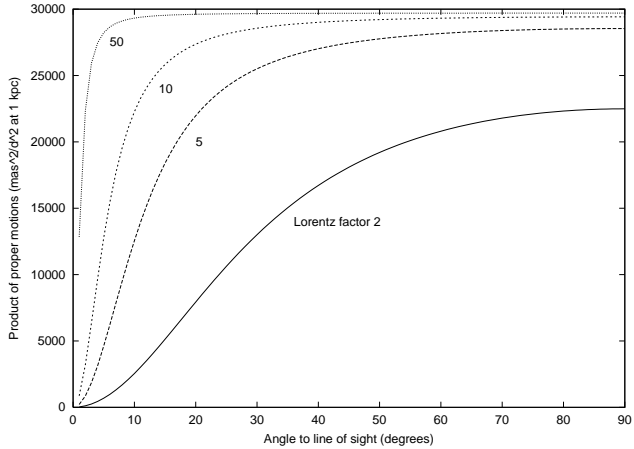, width=6cm, angle=270}}}
\caption{
Illustration of variation of the product $\mu_{\rm app} \mu_{\rm rec}$
as a function of angle to the line of sight for different Lorentz
factors. For the most relativistic jets, except at very small angles
to the line of sight the product remains constant. However, for only
moderately relativistic jets changes in the jet angle due to, for
example, precession, can cause significant changes in the
product. This may be a means to limit the Lorentz factor of the jets
if precession can be tracked.
}
\end{figure}

There is even a possibility to achieve the goal of limiting the
Lorentz factor, in the case of a jet whose angle to the line of sight
changes, for example due to precession. This can be seen from Fig 2,
where at lower Lorentz factors the $d_{\rm max}$ is quite a strong
function of angle, whereas it is not at all for the higher Lorentz
factors. This is illustrated in Fig 5, in which the product $\mu_{\rm
app} \mu_{\rm rec}$ is plotted for varying angles, for different
Lorentz factors. Apart from the smallest angles to the line of sight,
at which two-sided proper motions are anyway unlikely to be detected,
the most relativistic jets have an almost constant value of this
product, whereas the slower jets have a significantly varying
product. For example, a jet with a mean angle to the line of sight of
60 degrees, precessing with a half-opening angle of 20 degrees would
produce a $\sim 25$\% change in the product $\mu_{\rm app} \mu_{\rm
rec}$ over its precession period if it had an intrinsic Lorentz factor
of 2. If the jet has a Lorentz factor of five or more, the fractional
change in the product over the precession cycle is 5\% or less. This
approach, albeit almost certainly limited to galactic sources (due to
the necessity of tracking in time a periodic precession cycle)
presents an interesting possibility for limiting the Lorentz factors.

\section*{Conclusions}

In this paper the uses and limitations of relativistic jet proper
motions have been explored, under the assumption of intrinsically
symmetric ejection velocities. The main results derived are:

\begin{itemize}
\item{For the two galactic `microquasars' for which two-sided proper
motions have been measured, even the relatively small uncertainties in
the distances estimates result in an almost complete inability to
constrain the Lorentz and Doppler factors. Measurement of the flux
ratios of approaching and receding components does {\em not} provide
any additional constraints.}
\item{Exploring the general case, it is found that this will nearly
always be the situation -- i.e. that all relativistic jets will be
observed near the distance $d_{\rm max}$ at which $\beta \sim 1$ and
we will be unable to place an upper limit on the Lorentz factor of the
flow.  Conversely, this means that relativistic jet sources will
always be observed close to $d_{\rm max}$. This means that for AGN,
observations of two-sided proper motions will not allow accurate
measurement of the Lorentz factor of the jets, but {\em will} be
extremely important for calibration of the cosmological distance
scale, without requiring the observation of Doppler-shifted emission
lines.}
\item{It is shown that the product of the approaching and receding
proper motions varies significantly with angle to the line of sight
for jets which are only mildly relativistic, whereas for highly
relativistic jets the product is practically invariant. This opens up
the possibility of constraining the Lorentz factor of a precessing jet
by measurement of the product around the precession cycle.}
\end{itemize}

\subsection*{Caveats}

As stated above, everything calculated in this paper is only strictly
valid under the assumption of symmetric ejection events. Observations
of jets from the neutron star XRB Sco X-1 (Fomalont et al. 2001a,
2001b) have shown us that the resolved sites of radio emission may in
some cases simply be the regions of jet--ISM interaction and may not
reflect the underlying bulk velocity of the flow. This is even more
dramatically demonstrated by observations of large-scale decelerating
jets from the black hole transient XTE J1550-564 (Corbel et al. 2002;
Kaaret et al. 2003; Tomsick et al. 2003). As a result the application
of the results in this paper, e.g. estimating the distance $\sim
d_{\rm max}$ should, wherever possible, be based upon measurements as
early as possible in the flight of the ejecta. Finally, the observed
correlation between peak radio and X-ray fluxes from X-ray transients
(Fender \& Kuulkers 2001) would be destroyed if the Lorentz factor of
the radio emitting region were too large (unless the X-ray emission
were also beamed, which would however imply a huge selection effect on
observations of X-ray binaries) -- while current data may be too
sparse to constrain this at present, this may be the best approach for
limiting the Lorentz factors of jets from X-ray binaries in the
future.

\section*{Acknowledgements}

The author would like to thank Guy Pooley and Marc Ribo for comments
on this manuscript, and the anonymous referee for useful suggestions.

\end{document}